# A Superpixel Segmentation Based Technique for Multiple Sclerosis Lesion Detection

Saba Heidari Gheshlaghi[1], Amin Ranjbar[1], Amir Abolfazl Suratgar[1], Mohammad Bagher Menhaj[1], Fardin Faraji[2]

**Abstract—Automated and accurate classification of Magnetic Resonance Images (MRI) comes into account in medical analysis significantly, interpretation and improving the efficiency of healthcare in MS patients. Current study proposes a novel automatic classification system to distinguish Multiple Sclerosis's patients from healthy subjects by using their MR brain images. Herein, Support Vector Machine (SVM) as an effective classifier with Polynomial kernels is applied to have better performance in distinguishing specified decision classes. Furthermore, some methods come into account to have appropriate results such as discrete wavelet decomposition (DWT) which extracts local information from analyzing MR images, the superpixel segmentation to have automatic image partitioning which processed through principal components analyses method to deal with data dimensionality problem. The proposed method is tested with 10-fold cross validations method to check the final accuracy. Experiments using Amirkabir hospital dataset to classify MS lesion detection in Brain MR images depict that the current technique yields high performance outcome with an average accuracy up to 99%.**

## I. INTRODUCTION

**M**ultiple Sclerosis (MS) disease is considered as a kind of chronic diseases. MS always attacks the central nervous system and its white matter is affected through the patient's own immune system. As a result, MS is categorized as an auto-immune disease. Nerve fibers which covered by myelin, protects the nerves and helps them to conduct the electrical pulses. In MS patients, the myelin disappeared which is called Demyelination. [1] [2] Recent medical researches suggest that the genetic and environmental factors mainly cause MS. The fact that a monozygotic twin with a co-twin suffering from MS has a chance of 25% of developing the disease while the same chance for a non-twin sibling stands at only 3%, is an indicator of the role of genetic influence in developing MS. [3] Also, MS mostly affects adults in developed countries. Moreover, roughly 2.5 million people in the world suffer MS. Iran is considered as a country with high MS prevalence (51.52 per 100000) in the Middle East. [4] [5] Magnetic Resonance Imaging (MRI) is known as a useful method for monitoring and diagnosing MS disease. [6] [7] [8] MRI is based on the magnetic characteristics of the imaged tissue. This method is a painless, safe, fast, noninvasive imaging technique that yields images of the body structures, specifically capturing images within brain layers, and provides valuable information for medical diagnosis. [9] [10] [11] It is quite a tedious and challenging experience for an expert to analyze MR images due to the fact that there are complications and difficulties in terms of understanding and interpreting anatomical borders that are invisible almost in all images.

Through clinical routines, due to the abundant number of MR images, the regular exploration by a human expert always wastes valuable time for early diagnosis. Hence, automatically segmentation of brain images is always vital to substitute the manual segmentation. The advancing complexity of the MS lesions detection task comes from the various changes in the shape and location between patients, which turns the automatic segmentation to an intricate task. Here, a set of segmentation pipelines are suggested for MS lesions segmentation from human brain MRI.

In MR images, the most common sequences are T1-weighted, T2-weighted and Fluid Attenuated Inversion Recovery (FLAIR). Short TE and TR times produce T1-weighted images. T1 properties of tissue predominately determine the contrast

[1] Department of Electrical Engineering, Amirkabir University of Technology (Tehran Polytechnic), Tehran, Iran

[2] Neurology Department, Arak University of Medical Sciences, Arak, Iran

and brightness of the lesion. By doing so, Lesion (demyelination) and normal CSF are both dark. On the contrary, using longer TE and TR times produce T2-weighted images. In these images, Lesion and normal CSF are both bright. Also, Fluid Attenuated Inversion Recovery scans that are simply called "FLAIR' are another type of common sequence that is similar to a T2-weighted which TE and TR are longer makes lesion brighter than normal CSF. [12] This sequence brings high differentiation between CSF and an abnormality which makes corresponding analysis easier for further steps. In FLAIR images, the CFS is dark; the white matter is dark gray; fat is light, and demyelination leading to MS disease is bright in these images. Therefore, FLAIR is the most common sequence for lesions detection in MRI. Different reasons make automated lesion segmentation in MS a challenging task; the most important reasons are: (1) lesions are in different size and location, (2) lesion boundaries are not mostly well defined on FLAIR images, and (3) clinical quality FLAIR images may possess low resolution and noise problems. [13] [14]

During the last decades, MR imaging techniques have significantly contributed to the understanding and managing multiple sclerosis and have had a critical role in confirming the clinical diagnosis of MS. Various methods have been proposed for diagnosing MS lesions on brain MR image. Mostly, two common categories are applied in image segmentations; supervised and unsupervised; depending on the data source the appropriate one would be chosen. Abdullah Bassem (2012) developed a segmentation pipeline for automatic segmentation of MS lesions from brain MRI data. In that research, SVM derived by the textural information to distinguish between MS and Non-MS blocks. Furthermore, this research introduces the concept that uses multi-sectional views to produce a verified segmentation. This method showed results with 68% accuracy. [15] Van Leemput Koen, et al. (2001) used a stochastic model which is called Markov Random Field on the multispectral MR images and it was discussed the outlier detection with contextual information. According to the results, there was a high total lesion load correlation compared with other methods. [16] Heidari Gheshlaghi, S. et al. (2018) proposed a fuzzy method for diagnosing MS disease from brain MR images. In this research, well-known image processing methods such as edge detection were applied. By changing and reforming fuzzy c-means clustering algorithms, and applying canny contraction principles, the relationship between MS lesions and edge detection was established. [17] Mechrez et al. (2016) present a method for lesions segmentation. This paper works by using similarities amongst multichannel areas. The areas databank is provided to learn labeld images. They use similar patches for the testing image. Finally, based on the initial segmentation map, an iterative patch-based label is performed to ensure the spatial uniformity of distinguished lesions. In addition, applying a more efficient database characterization method along with advanced metric learning can directly contribute to the enhancement and improvement of lesion detection results. [18] Moreover, Roy, Snehashis, et al. (۲۰۱۸) propose a fully Convolutional Neural Networks for a procedural MS lesion detection through MRI data. They introduced Cascade neural network which contains two convolutional pathways. The first pathway includes multiple parallel convolutional filter banks that cater to a wide variety of MRI modalities, and in the second pathway, the concatenation of the output of the first pathway is done. The output produces a membership function for MS lesions that may be the threshold to obtain a binary segmentation. This method showed results with 90.48% accuracy. [19]

## II. Processing

Raw images are not suitable for analysis due to having many artifacts such as intensity inhomogeneity, extracranial tissues, and noises. These unwanted data reduce the accuracy and efficiency of segmentation. Therefore, suitable pre-processing techniques must be used to improve the quality of the image for further steps. Literature study shows several pre-processing and feature extraction methods for MR brain image analysis. In this part, pre-processing and segmentation methods will be discussed in detail.

### A. Brain Extraction

For having a valuable and accurate MR image, noise reduction and skull extraction are crucial. There are many different technics for skull extraction [20] [21] although this step is a significant part, using an efficient method is essential. Brain Extraction Tool (BET) [22] [23] [24], Brain Surface Extraction (BSE) [25], Watershed Algorithm (WAT) [26], Hybrid Watershed Algorithm (HWA) [27] and Skull Stripping using Graph Cuts [28] are some of the famous approaches for brain extracting in MR images. MR brain images have different limitations such as resolution, noise, low contrast and geometric deformations which are the most significant limitations for the development of an efficient brain segmentation algorithm. [29] [30] [31]

There are some non-brain tissues in MR brain images, and these tissues have no valuable data for analysis, so it is necessary to remove these tissues for further processing. The Brain Extraction Tool (BET) is one of the well-known brain splitting tools which removes non-brain tissues in MR brain images. This method is simple and one of the robust brain extraction tools. [32] BET tool is publicly available in the FSL repository [33], and it follows the following steps to extract desired area of the brain:

- Histogram-based threshold estimation
- Binarization of the image

- Finding the center of the image
- Initializing the triangular tessellated sphere surface
- Surface deformation
- Extraction of brain boundary

In our research, we used the automated brain extraction tool (Smith, 2002). Also, image binarization which was described in the BET method is presented in equation 1 where th represents the threshold. [34] [35] [36]

$$Bin(x,y) = \begin{cases} 1 & if & f(x,y) > th \\ 0 & if & O.w \end{cases} \qquad (1)$$

## B. Superpixel segmentation

Superpixels have become increasingly popular in recent years. Ren and Malik (2003) first introduced superpixel. [37] Superpixels group pixels into regions that can be used to replace the rigid structure of the pixel grid in images. This method detects repetitive patterns in the image and greatly reduces the calculation complexity and also the processing time. Simple Linear Iterative Clustering Superpixel (SLIC) segmentation is becoming more common, and similar to the K-Means algorithm, pixels are clustered in accordance to their color similarity. [38] [39] In this research, many pixels must be examined, and it may cause errors and turn out to be time-consuming. In SLIC algorithm, the number of K initial cluster centers is defined randomly. Next, the algorithm proceeds to assign and then update parameters in an iterative manner. In this step, according to a similarity measurement, each pixel belongs to the closest cluster. In the next step, the clusters would be updated in terms of their member pixels which have been assigned during the updating step; the process would continue until convergence is reached. [40] [41] [42]

It is noticeable that this method gets structured on the basis of "lab" color space. Moreover, it is not possible to use regular Euclidean distance here for solving the problem; hence, a new distance measuring method is introduced. Primarily, the gray scale images are converted into "lab" space. As it was mentioned, in superpixel algorithm, K has been introduced as the desired input number of approximately equally-sized superpixels or clusters. The estimated size for each superpixel is calculated as $\frac{N}{K}$ pixels for each cluster where "N" stands for total number of pixels and in following "S" shows the superpixel centers. [43] [44] [45]

At the beginning of superpixel algorithm, the number of K superpixel centers is chosen $C_k = [l_k, a_k, b_k, x_k, y_k]^T$ with $K = [1, k]$ at any fixed interval "S". As the spatial scope of this parcelled region is approximately $S^2$, it can be supposed that pixels associated with this cluster center locates in the area of $2S \times 2S$ around the center. For measuring distance, "Ds" is:

$$d_{lab} = \sqrt{(l_k - l_i)^2 + (a_k - a_i)^2 + (b_k - b_i)^2}$$
$$d_{xy} = \sqrt{(x_k - x_i)^2 + (y_k - y_i)^2} \qquad (2)$$
$$D_s = d_{lab} + \frac{m}{s} d_{xy}$$

The summation over all lab distances which is defined as "Ds", has to be normalized in x-y plane. Variable m is introduced as a weighting factor to make a balance between color and spatial differences which its range is [1, 20]. The number of desired superpixels is introduced as k. In this project, we use m=5 and k=500.

Where:

$$G(x,y) = \parallel I(x+1,y) - I(x-1,y) \parallel^2 + \parallel I(x,y+1) - I(x,y-1) \parallel^2 \qquad (3)$$

I(x,y) is converted version of the original image to "lab" space, and ∥.∥ depicts the L2 norm. Each pixel of the image is connected with the nearest cluster containing this pixel. After finishing calculation on all the pixels, by averaging labxy vector of all the pixels belonging to the cluster, a new cluster is considered. Then, the process that includes associating pixels with the nearest cluster center and also re-computing the cluster center would be repeated until convergence is reached.

## C. Discrete Wavelet Transform

Fourier's representation functions as a superposition of sines and cosines and represents the frequency domain of the original data. The main flaw of the Fourier transform is that it fails to represent where the signal has several discontinuities and sharp spikes. [46] [47] For solving this problem, Wavelets have been introduced for time-frequency analysis. Grossmann and Morlet first introduced wavelets. Wavelets are mathematical functions that divide data or signals into different frequency components. Discrete Wavelet Transforms (DWT) is the most famous transformation technique adopted for image compression. In DWT, a digital signal is analyzed in terms of time and frequency content which profit from the filtering methods. [48] [49]

Wavelet packet coefficients of a finite energy function f(t) with the wavelet packet functions $W_{j,k,n}$ presented in the following equations:

$$C_{j,k,n} = \int_{-\infty}^{\infty} f(t) W_{j,k,n} dt \qquad (4)$$

$$W_{j,k,n}(t) = 2^{(-j/2)}W_n(2^{-j}t - k) \quad n \in \mathbb{N}, \quad j,k \in \mathbb{Z} \tag{5}$$

Where $n$, $j$, $k$ denote modulation, resolution and translation index respectively. Besides, the wavelet function effects are represented by the coefficients in the following:

$$W_{2n}(t) = \sqrt{2} \sum_{-\infty}^{\infty} h_0[k] W_n(2t - k) \tag{6}$$

$$W_{2n+1}(t) = \sqrt{2} \sum_{-\infty}^{\infty} g_0[k] W_n(2t - k) \tag{7}$$

In previous equations, $W_0(x) = \phi(x)$ and $W_1(x) = \psi(x)$, additionally, $h_0$ and $g_0$ indicate lowpass and highpass filters obtained by proper wavelet function.

The DWT is applied to MR images and extracts approximated and detailed version of images hierarchically. As decomposition level increases more stable approximation coefficients are obtained. In our method, we use 2 level DWT, and the design simulated in MATLAB.

### D. Texture Extraction Techniques

The statistical information of each region of interest i.e. the superpixels used to describe relevant features distributed in the clusters. In each superpixel, four features including mean, variance, Skewness, and kurtosis are evaluated over the analytical image $f(x,y)$. The function f(x,y) can have any value $k = 0, 1, \ldots, L-1$ where $z_k$ is the total number of intensity levels corresponding to the image. Some occurrences of $z_k$ in the $M \times N$ image is given by $n_k$. Each of these momentums is calculated for each superpixel.

Estimated probability density regards to intensity level image is defined as $p(z_k)$:

$$p(z_k) = \frac{n_k}{MN} \tag{8}$$

The statistical textural features are calculated as following:

Mean:
$$\mu = \sum_{k=0}^{L-1} z_k \, p(z_k) \tag{9}$$

Variance:
$$\sigma^2 = \sum_{k=0}^{L-1} (z_k - \mu)^2 \, p(z_k) \tag{10}$$

Skewness:
$$\mu_3 = \frac{1}{\sigma^3} \sum_{k=0}^{L-1} (z_k - \mu)^3 \, p(z_k) \tag{11}$$

Kurtosis:
$$\mu_4 = \frac{1}{\sigma^4} \sum_{k=0}^{L-1} (z_k - \mu)^4 \, p(z_k) - 3 \tag{12}$$

### E. Principal Component Analysis

Principal Component Analysis (PCA) uses an orthogonal transformation for converting correlated variables into uncorrelated variables by using mathematical methods. The idea was initially conceived by Pearson (1901) and independently developed by Hotelling (1933); hence this technique is known as Hotelling transform. [50] PCA can decrease any data dimension procedurally to a more informative set of variables. PCA is a member of linear transforms based on the statistical techniques. [51] This method introduces an excellent tool to compress the data and diminish the dimensions in data analysis and pattern recognition area. [52] [53] [54] [55] [56]] PCA transforms $n$ input vectors which formed as $x = [x_1, x_2, \ldots x_n]^T$ into a final vector $y$ according to:

$$y = A(x - m_x) \tag{13}$$

The vector $m_x$ in the equation above is the vector of mean values over all variables and is defined as follows:

$$m_x = E\{x\} = \frac{1}{k} \sum_{k=1}^{k} x_k \tag{14}$$

The matrix A in Eq. (12) is calculated by the covariance matrix $C_x$. Rows in the $A$ matrix are generated from the eigenvectors $e$ of $C_x$ order according to analogous eigenvalues in descendant order. The evaluation of the $C_x$ is:

$$C_x = E\{(x - m_x)(x - m_x)^T\} = \frac{1}{k} \sum_{k=1}^{k} x_k \, x_k^T - m_x m_x^T \tag{15}$$

The size of $C_x$ is $n \times n$ because the vector $x$ of input variables is $n$-dimensional. The elements are placed in the main diagonal of $C_x(i,i)$ are called the variances of $x$, and the rest reflect the covariance values distinguished input variables. [57]

$$C_x(i,i) = E\{(x_i - m_{x_i})^2\} \tag{16}$$

Due to the orthonormal property, inversion of A comes below:

$$x = A^T y + m_x \tag{17}$$

In this research, 16 features were extracted originally. While, by applying PCA, the number of features reduced to 10.

## III. CLASSIFICATION

Data classification acts as one of the highly important phases of machine learning which, depending on the readiness of the training data, can follow a supervised, unsupervised or semi-supervised mode. [58] [59] In this research, After Feature Extraction and Feature Reduction, we fed the data into the supervised classifier. Support Vector Machines (SVM), according to Zhang, Yudong, et al. (2015), Abdullah N. et al. (2011), Shigeo, A. (2005) enjoy a number of advantages the most important of which include high accuracy even with high dimensionality at input data while the final performance still are kept. [60] [61] [62] [63] Therefore, these properties make the SVM an appropriate way for MS lesion detection in MR images.

### A. Support Vector Machine

Support Vector Machine which is mostly known as SVM was presented by Vapnik (1996) for the first time which has been developed based on an idea that creates a hyperplane between data sets in order to find out which class a certain data set belongs to.

This part of research uses SVM to classify the images, and they are divided into two groups, normal and abnormal conditions. The process contains two components, which are training part and a testing part. [64] [65] [66] [67] In what follows, x stands for a vector containing components $x_i$. In a dataset, the $i^{th}$ vector will be denoted by the notation $x_i$ where $y_i$ is the label related to $x_i$.

The discriminant function in the space F is:

$$f(x) = w^T \phi(x) + b \qquad (18)$$

While applying the mapping, the approach of explicit computation of non-linear features seems not to be scaling well with the number of input features. As a result, it leads to rise in memory usage for processing the features at the same time, and it reduces the computing time. Kernel methods have their own strategy of solving the non-linear issue. [68] Assume the weight vector can be definied as a linear combination of the training data, i.e.

$$w = \sum_{i=1}^{n} \alpha_i x_i \qquad (19)$$

Then:

$$f(x) = \sum_{i=1}^{n} \alpha_i x_i^T x + b \qquad (20)$$

Then, we have:

$$f(x) = \sum_{i=1}^{n} \alpha_i \phi(x_i^T) \phi(x) + b \qquad (21)$$

The $\alpha_i$ in last equation, can be high dimensional and accordingly reformed to the kernel function $k(x, x')$ as following:

$$K(x, x') = \phi(x)^T \phi(x') \qquad (22)$$

So, finally we have:

$$f(x) = \sum_{i=1}^{n} \alpha_i k(x, x_i) + b \qquad (23)$$

Kernalization algorithm is essential for having useful training, and many algorithm in the field of machine learning uses kernels like the perceptron algorithm, ridge regression, and SVMs. [69] [70] The common basic kernel functions are:

Linear Kernel:

$$K(x_i, x_j) = x_i^T x_j \qquad (24)$$

Polynomial:

$$K(x_i, x_j) = (\gamma x_i^T x_j + r)^d \qquad (25)$$

Radial basis function (RBF):

$$K(x_i, x_j) = \exp(\gamma \| x_i - x_j \|^2); \ \gamma > 0 \qquad (26)$$

Sigmoid function:

$$K(x_i, x_j) = \tanh(\gamma x_i^T x_j + r) \qquad (27)$$

Where $\gamma$, r and d are kernel parameters.

### B. Cross-Validation

Cross-Validation is a common method which is applied for running comparisons amongst learning algorithms. The data are separated into two sections; one section is used for learning or training the model, and the other one issued for validating the model. The most famous form of cross-validation is k-fold cross-validation method. There are other forms of cross-validation such as Hold-Out Cross-Validation method. [71] The following table shows the differences between these two popular cross-validation methods:

| Validation method | Positive points | Negative points |
|---|---|---|
| Hold-out | Independence between training and test data | data training and testing reduced; Large variance |
| k-fold | Good accuracy performance estimation | estimation samples are small; overlapped training data |

*Table 1: popular cross-validation methods*

## IV. RESULTS

In this research, the pipeline is established based on statistical information as a feature vector and SVM ensemble in the MS lesions detection task. Datasets of 70 cases were come into account to verify the anticipated technique. The dataset consists of 35 MR images with MS lesion and 35 MR images from healthy subjects. The sources of these datasets are from the Amirkabir Hospital, Arak, Iran.

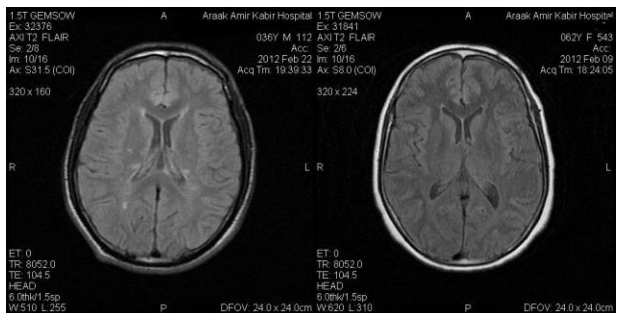

*Figure 1: Raw MR images[3]*

In the training phase, first, removing unwanted data and noise reduction from MR images are needed. This step is an essential part because if we do not remove the skull correctly from the raw images; all the next processes will be affected. In this step, BET boundary removal and binarization ensemble were applied to have the most valuable area of brain.

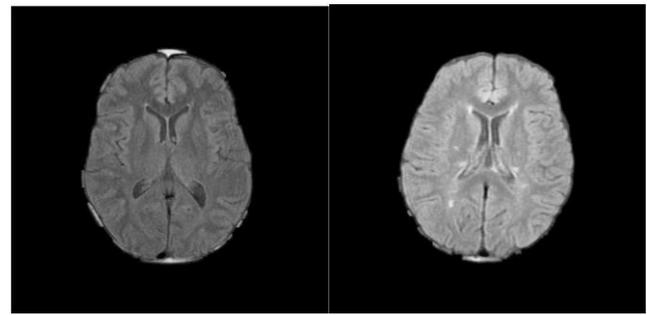

*Figure 2: MR image after applying Brain Extraction Tools*

For next step, we applied superpixel segmentation method on the image dataset to partition all images and detecting desired lesion area efficiently. Furthermore, because superpixel segments all the images and borders do not have valuable information, for this step, these borders have been removed from MR images to reduce the calculation and increase the performance. After Skull extraction, there are still some parts that may cause some errors in our future classification. For solving this problem, after applying superpixel method and removing borders, we remove the white parts that are neighbors with our borders to overcome these unwelcome regions. This step and further processing are illustrated in the following figures. At first step, the target amount for desired superpixels is 500, and the weighting factor asumes to be 5. Furthermore, borders are removed. The results are shown below:

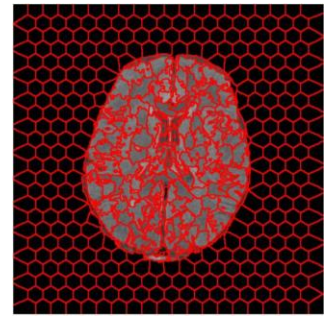

*Figure 3: MR images after applying superpixel method*

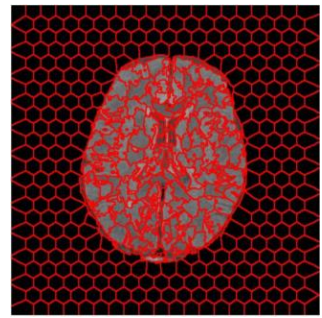

*Figure 4: MR images after applying superpixel method and removing borders*

---

[3] Due to the patient's rights, the patient's name is deleted.

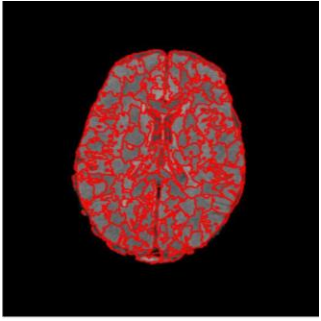

*Figure 5: MR images after applying superpixel method and removing borders*

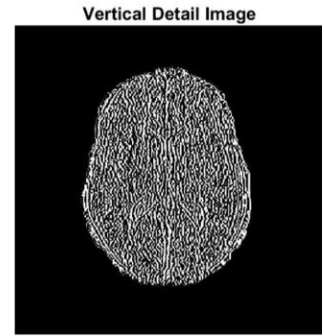

*Figure 9: Vertical details decomposition of input image via one level 2-D DWT*

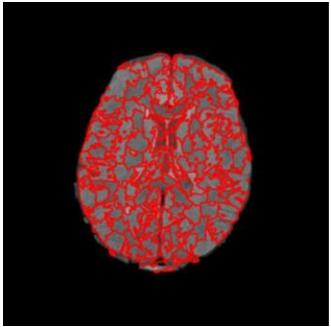

*Figure 6: MR images after applying superpixel and removing unwanted superpixels*

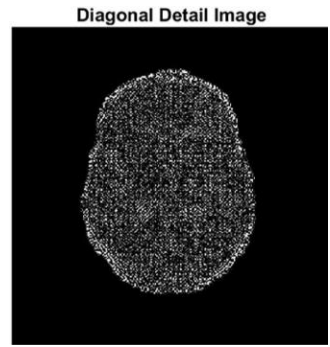

*Figure 10: Diagonal details decomposition of input image via one level 2-D DWT*

For the next step, Discrete 2-D wavelet transform is applied.

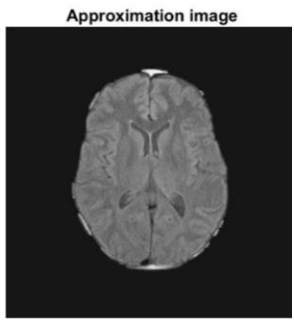

*Figure 7: Approximation decomposition of input image via one level 2-D DWT*

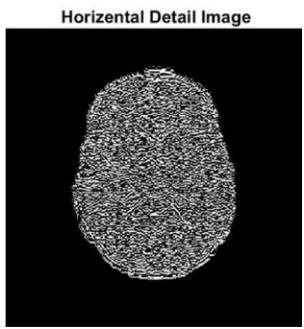

*Figure 8: Horizontal details decomposition of input image via one level 2-D DWT*

For next step, four above-mentioned statistical features are calculated within the area of each superpixel. These features are the most common features for investigations of MR images. After the manipulations which were mentioned above, feature extraction is done entirely, and at the last step we go through training SVM classifier using the cross validated dataset as described in previous section. Different kernel function including RBF, polynomial and quadratic function are evaluated to achieve better performance. Corresponding results are shown in Table2.

| Kernel | Cross-Validation Method | Accuracy |
|---|---|---|
| RBF Function | 10-fold | 0.9965% |
| RBF Function | holdout | 0.9957% |
| Polynomial Function | 10-fold | 0.9991% |
| Polynomial Function | holdout | 0.9968% |
| Quadratic Function | 10-fold | 0.9981% |
| Quadratic Function | holdout | 0.9970% |

*Table 2: Classification results using test images for different kernel functions*

## V. CONCLUSION

Multiple Sclerosis occurs when myelin is attacked by the immune system. Myelin covers human fiber in order to protect the nerves and help them send messages efficiently. Without this protective layer, nerves are vulnerable to be damaged which would ultimately undermine transferring signals between brain and body. Doctors review a number of critical factors including the medical history, physical checks, neurological examinations, particularly those via MR imaging method in order to diagnose the disease. MR brain images continue to be a valuable and efficient tool for diagnosing Multiple Sclerosis disease. Due to the sensitivity of MR images, it is a difficult task to clinically diagnose the lesions, and it takes a long time to do so. So applying an automatic method for diagnosing the abnormal lesion would equal taking a critical step towards improving both the speed and accuracy of diagnosis.

In this research, a lesion detection framework was developed for a procedural classification technique in MS lesions detection regards to MR brain images. This pipeline has been developed to use the statistical information and discrete wavelet transform to extract highly relevant features and SVM classifier to discriminate the normal and abnormal regions. Furthermore, to reduce calculation complexity and increasing the performance of the proposed method, superpixel segmentation algorithm has been applied. In this research, real datasets with 70 cases were used. Also, in this dataset, 35 of MR images are from MS patients with a brain lesion, and 35 MR images are from healthy subjects. The sources of these datasets are from the Amirkabir Hospital, Arak, Iran. Besides FLAIR images have been used for segmentation.

For summarizing the algorithm, the initial step includes pre-processing methods for noise reduction and removing unwanted regions from MR brain images. At this step, which is a critical and vital step to proceed, we used BET application and binarization. Furthermore, we applied superpixel segmentation method for segmentation and removing the pixels that lack information and may cause an error in our classification. This step increases our performance and accuracy. Also, for feature extraction, discrete wavelet transform along with statistical features have been applied, and PCA is used to reduce the features that have no valuable data for us and may create problems in our final results. Finally by applying popular cross validations methods, the accuracy were checked. This new classification method was able to distinguish the normal and abnormal regions in MR brain images with an overall efficiency of 99%.

## REFERENCES

[1] Polman, C. H., Reingold, S. C., Banwell, B., Clanet, M., Cohen, J. A., Filippi, M., ... & Lublin, F. D. , "Diagnostic criteria for multiple sclerosis: 2010 revisions," *Annals of Neurology,* vol. 69, no. 2, pp. 292-302, 2011.

[2] Ghribi, O., Sellami, L., Slima, M. B., Mhiri, C., Dammak, M., & Hamida, A. B., "Multiple sclerosis exploration based on automatic MRI modalities segmentation approach with advanced volumetric evaluations for essential feature extraction," *Biomedical Signal Processing and Control,* vol. 40, pp. 473-487, 2018.

[3] Willer, C. J., Dyment, D. A., Risch, N. J., Sadovnick, A. D., Ebers, G. C., & Canadian Collaborative Study Group., "Twin concordance and sibling recurrence rates in Multiple sclerosis," *Proceedings of the National Academy of Sciences,* vol. 100, no. 2, pp. 12877-12882, 2003.

[4] Eskandarieh, S., Heydarpour, P., Elhami, S. R., & Sahraian, M. A., "Prevalence and Incidence of Multiple Sclerosis in Tehran, Iran," *Iranian Journal of public health,* vol. 46, no. 5, p. 699, 2017.

[5] Rahimi, F., Rasekh, H. R., Abbasian, E., Peiravian, F., Etemadifar, M., Ashtari, F., ... & Amirsadri, M. R, "Patient preferences for Interferon-beta in Iran: A discrete choice experiment," *PLOS ONE,* 2018.

[6] Miller, D. H., Grossman, R. I., Reingold, S. C., & McFarland, H. F., "The role of magnetic resonance techniques in understanding and managing multiple sclerosis," *Brain: a journal of neurology,* vol. 121, no. 1, p. 3–24, 1998.

[7] McDonald, W. I, "Diagnosis of multiple sclerosis," *British Medical Journal(BMJ),* vol. 299, no. 6700, pp. 635-637, 1989.

[8] Thompson, A. J., Banwell, B. L., Barkhof, F., Carroll, W. M., Coetzee, T., Comi, G., ... & Fujihara, K. , "Diagnosis of multiple sclerosis: 2017 revisions of the McDonald criteria," *The Lancet Neurology,* pp. 1-9, 2017.

[9] Bernheim, K. A. , "Functional and structural magnetic resonance imaging of humans and macaques," California Institute of Technology, 2004.

[10] Boudraa, A. O., Dehak, S. M. R., Zhu, Y. M., Pachai, C., Bao, Y. G., & Grimaud, J., "Automated segmentation of multiple sclerosis lesions in multispectral MR imaging using fuzzy clustering," *Computers in biology and medicine,* vol. 30, no. 1, pp. 23-40, 2000.

[11] Nayak, D. R., Dash, R., & Majhi, B., "An improved extreme learning machine for pathological brain detection," *Region 10 Conference, TENCON 2017-2017 IEEE,* pp. 13-18, 2017.

[12] Khayati, R., Vafadust, M., Towhidkhah, F., & Nabavi, M., "Fully automatic segmentation of multiple sclerosis lesions in brain MR FLAIR images using adaptive mixtures method and Markov random field model," *omputers in biology and medicine,* vol. 38, no. 3, pp. 379-390, 2008.

[13] Carass, A., Roy, S., Jog, A., Cuzzocreo, J. L., Magrath, E., Gherman, A., ... & Cardoso, M. J. , "Longitudinal multiple sclerosis lesion segmentation: Resource and challenge," *NeuroImage ,* vol. 148, pp. 77-102, 2017.

[14] Alshayeji, M. H., Al-Rousan, M. A., Ellethy, H., & Abed, S. E., "An efficient multiple sclerosis segmentation and detection system using neural networks," *Computers & Electrical Engineering,* vol. 71, pp. 191-205, 2018.

[15] B. A. Abdullah, "Segmentation of Multiple Sclerosis Lesions in Brain MRI," University of Miami, 2012.

[16] Van Leemput, K., Maes, F., Vandermeulen, D., Colchester, A., & Suetens, P. , "Automated segmentation of multiple sclerosis lesions by model outlier detection," *IEEE transactions on medical imaging,* vol. 20, no. 8, pp. 677-688, 2001.

[17] Heidari Gheshlaghi, S., Madani, A., Suratgar, A., & Faraji, F. , "Segmentation of Multiple Sclerosis lesion in brain MR images using Fuzzy C-Means," *International Journal of Artificial Intelligence and Applications (IJAIA),* vol. 9, no. 2, pp. 37-45, 2018.

[18] Mechrez, R., Goldberger, J., & Greenspan, H. , "Patch-based segmentation with spatial consistency: application to MS lesions in brain MRI," *Journal of Biomedical Imaging,* vol. 2016, no. 3, 2016.

[19] Roy, S., Butman, J. A., Reich, D. S., Calabresi, P. A., & Pham, D. L., "Multiple Sclerosis Lesion Segmentation from Brain MRI via Fully Convolutional Neural Networks," *arXiv preprint arXiv:1803.09172,* 2018.

[20] Clarke, L. P., Velthuizen, R. P., Camacho, M. A., Heine, J. J., Vaidyanathan, M., Hall, L. O., ... & Silbiger, M. L., "MRI segmentation: methods and applications," *Magn Reson,* vol. 13, no. 3, pp. 343-368, 1995.

[21] Ali, S. M., & Maher, A., "Identifying multiple sclerosis lesions in MR images using image processing techniques," in *Multidisciplinary in IT and Communication Science and Applications (AIC-MITCSA), Al-Sadeq International Conference,* 2016.

[22] Popescu, V., Battaglini, M., Hoogstrate, W. S., Verfaillie, S. C., Sluimer, I. C., van Schijndel, R. A., ... & Barkhof, F., "Optimizing parameter choice for FSL-Brain Extraction Tool (BET) on 3D T1 images in multiple sclerosis," *Neuroimage,* vol. 61, no. 4, pp. 1484-1494, 2012.

[23] Eskildsen, S. F., Coupé, P., Fonov, V., Manjón, J. V., Leung, K. K., Guizard, N., ... & Alzheimer's Disease Neuroimaging Initiative., "BEST: brain extraction based on nonlocal segmentation technique.," *NeuroImage,* vol. 59, no. 3, pp. 2362-2373, 2012.

[24] Rex, D. E., Shattuck, D. W., Woods, R. P., Narr, K. L., Luders, E., Rehm, K., ... & Toga, A. W., "meta-algorithm for brain extraction in MRI," *NeuroImage,* vol. 23, no. 2, pp. 625-637, 2004.

[25] Shattuck, D. W., Sandor-Leahy, S. R., Schaper, K. A., Rottenberg, D. A., & Leahy, R. M. , "Magnetic Resonance Image Tissue Classification using a Partial Volume Model," *Neuroimage,* vol. 13, no. 5, pp. 856-876, 2001.

[26] Hahn, H. K., & Peitgen, H. O., "The Skull Stripping Problem in MRI Solved by Single 3D Watershed Transform," in *International Conference on Medical Image Computing and Computer-Assisted Intervention,* Springer, Berlin, Heidelberg, 2000.

[27] Ségonne, F., Dale, A. M., Busa, E., Glessner, M., Salat, D., Hahn, H. K., & Fischl, B. , "A Hybrid Approach to the Skull Stripping Problem in MRI," *Neuroimage,* vol. 22, no. 3, pp. 1060-1075, 2004.

[28] Sadananthan, S. A., Zheng, W., Chee, M. W., & Zagorodnov, V. , "Skull Stripping using Graph Cuts," *Neuroimage,* vol. 49, no. 1, pp. 225-239, 2010.

[29] Somasundaram, K., & Kalavathi, P., "A hybrid method for automatic skull stripping of magnetic resonance images (MRI) of human head scans," *Computing Communication and Networking Technologies (ICCCNT), 2010 International Conference o,* pp. 1-5, 2010.

[30] Roy, S., Knutsen, A., Korotcov, A., Bosomtwi, A., Dardzinski, B., Butman, J. A., & Pham, D. L. , "A deep learning framework for brain extraction in humans and animals with traumatic brain injury," *Biomedical Imaging (ISBI 2018), 2018 IEEE 15th International Symposium on.,* pp. 687-691, 2018.

[31] Roy, S., Bhattacharyya, D., Bandyopadhyay, S. K., & Kim, T. H. , "An effective method for computerized prediction and segmentation of multiple sclerosis lesions in brain MRI," *Computer methods and programs in biomedicine,* vol. 140, pp. 307-320, 2017.

[32] P. Kalavathi, "Improved Brian Extraction tool using marker-controlled watershed segmentation," *IOSR Journal of computer Engineering,* vol. 16, no. 5, pp. 106-111, 2014.

[33] S. Smith, "Fast robust automated brain extraction," *Human Brain Mapping,* vol. 17, no. 3, pp. 143-155, November 2002.

[34] Denitto, M., Farinelli, A., Figueiredo, M. A., & Bicego, M. , "A biclustering approach based on factor graphs and max-sum algorithm," *Pattern Recognition,* vol. 62, pp. 114-124, 2017.

[35] Somasundaram, K., & Kalavathi, P., "A novel skull stripping technique for T1-weighted MRI human head Scans," in *Proceedings of the Eighth Indian Conference on Computer Vision, Graphics and Image Processing,* 2012.

[36] Alansary, A., Ismail, M., Soliman, A., Khalifa, F., Nitzken, M., Elnakib, A., ... & Zurada, J. M., "Infant Brain Extraction in T1-weighted MR Images using BET and Refinement using LCDG and MGRF Models," *IEEE journal of biomedical and health informatics,* vol. 20, no. 3, pp. 925-935, 2016.

[37] Ren, X., & Malik, J., "Learning a classification model for segmentation," in *Ninth IEEE International Conference on Computer Vision,* Nice, France, 2003.

[38] Achanta, R., Shaji, A., Smith, K., Lucchi, A., Fua, P., & Süsstrunk, S., "SLIC Superpixels," *EPFL Technical,* 2010.

[39] Xie, X., Xie, G., & Xu, X., "High precision image segmentation algorithm using SLIC and neighborhood rough set," *Multimedia Tools and Applications,* pp. 1-19, 2018.

[40] Sun, Z., & Chi, M., "Superpixel-Based Active Learning For The Classification Of Hyperspectral Images," *Hyperspectral Image and Signal Processing: Evolution in Remote Sensing (WHISPERS),* pp. 1-4, 2015.

[41] Hong, S. M., & Ho, Y. S., "Depth map refinement using superpixel label information," in *Signal and Information Processing Association Annual Summit and Conference (APSIPA,* Asia-Pacific, 2016.

[42] Csillik, O., "Fast segmentation and classification of very high resolution remote sensing data using SLIC superpixels," *Remote Sensing,* vol. 9, no. 3, p. 243, 2017.

[43] Wei, X., Yang, Q., Gong, Y., Ahuja, N., & Yang, M. H. , "Superpixel Hierarchy," arXiv preprint arXiv:1605.06325, 2016.

[44] Zhang, Y., Li, X., Gao, X., & Zhang, C., "A Simple Algorithm of Superpixel Segmentation With Boundry Constraint," *IEEE Transactions on Circuits and Systems for Video Technology,* vol. 27, no. 7, pp. 1-13, 2017.

[45] Sun, Z., & Chi, M., "Superpixel-based active learning for the classification of hyperspectral images," *Hyperspectral Image and Signal Processing: Evolution in Remote Sensing (WHISPERS), 2015 7th Workshop on. IEEE,* pp. 1-4, 2015.

[46] A. Graps, "An introduction to wavelets," *IEEE computational science and engineering,* vol. 2, no. 2, pp. 50-61, 1995.

[47] Lewis, A. S., & Knowles, G. , "Image compression using the 2-D wavelet transform," *IEEE Transactions on image Processing,* vol. 1, no. 2, pp. 244-250, 1992.

[48] Gupta, D., & Choubey, S., "Discrete Wavelet Transform for Image Processing," *International Journal of Emerging Technology and Advanced Engineering,* vol. 4, no. 3, pp. 598-602, 2015.

[49] Wang, S., Zhang, Y., Dong, Z., Du, S., Ji, G., Yan, J., ... & Phillips, P., "Feed-forward neural network optimized by hybridization of PSO and ABC for abnormal brain detection," *International Journal of Imaging Systems and Technology,* vol. 25, no. 2, pp. 153-164, 2015.

[50] Nixon, M. S., & Aguado, A. S., Feature extraction & image processing for computer vision, Academic Press, 2014.

[51] Everitt, B. S., & Dunn, G., "Principal components analysis," *Applied multivariate data analysis,* pp. 48-73, 2001.

[52] Mudrova, M., & Procházka, A., "Principle Component Aalysis In Image Processing," in *In Proceedings of the MATLAB Technical Computing Conference,* Prague, 2005.

[53] Barrett, H. H., & Myers, K. J. , "Foundations of Image Science," *John Wiley & Sons, New Jersey,* vol. 87, no. 1, p. 93, 2004.

[54] Smith, L. I. , "A tutorial on Principal Components Analysis," 2002.

[55] Dunteman, G. H. , Principal components analysis, Sage: Sage Publication, 1989.

[56] Shui, W., Zhou, M., Maddock, S., He, T., Wang, X., & Deng, Q. , "A PCA-Based method for determining craniofacial relationship and sexual dimorphism of facial shapes," *Computers in biology and medicine,* vol. 90, pp. 33-49, 2017.

[57] Xu, D., & Wang, Y., "An automated feature extraction and emboli detection system based on the PCA and fuzzy sets," *Computers in Biology and Medicine,* vol. 37, no. 6, pp. 861-871, 2007.

[58] Mayer, R., Simone, C. B., Skinner, W., Turkbey, B., & Choykey, P., "Pilot study for supervised target detection applied to spatially registered multiparametric MRI in order to non-invasively score prostate cancer," *Computers in biology and medicine,* vol. 94, pp. 65-73, 2018.

[59] Portela, N. M., Cavalcanti, G. D., & Ren, T. I., "Semi-supervised clustering for MR brain image segmentation," *Expert Systems with Applications,* vol. 41, no. 4, pp. 492-1497, 2014.

[60] Zhang, Y., Dong, Z., Liu, A., Wang, S., Ji, G., Zhang, Z., & Yang, J. , "Magnetic resonance brain image classification via stationary wavelet transform and generalized eigenvalue proximal support vector machine," *Journal of Medical Imaging and Health Informatics,* vol. 5, no. 7, pp. 1395-1403, 2015.

[61] Abdullah, N., Ngah, U. K., & Aziz, S. A., "Image classification of brain MRI using support vector machine," in *IEEE International Conference,* 2011.

[62] Shigeo, A., "Support vector machines for pattern classification," *Advances in Pattern Recognition,* vol. 2, 2005.

[63] Praveen, G. B., Agrawal, A., Sundaram, P., & Sardesai, S., "Ischemic stroke lesion segmentation using stacked sparse autoencoder," *Computers in biology and medicine,* 2018.

[64] Raghtate, G. S., & Salankar, S. S., "Automatic Brain MRI Classification Using Modified Ant Colony System and Neural Network Classifier," in *Computational Intelligence and Communication Networks (CICN) 2015 International Conference,* 2015.

[65] Tang, X., Zeng, W., Shi, Y., & Zhao, L., "Brain activation detection by modified neighborhood one-class SVM on fMRI data," *Biomedical Signal Processing and Control,* vol. 39, pp. 448-458, 2018.

[66] Abdullah, N., Ngah, U. K., & Aziz, S. A, "Image classification of brain MRI using support vector machine," in *Imaging Systems and Techniques (IST), 2011 IEEE International Conference on,* 2011.

[67] Trigui, R., Mitéran, J., Walker, P. M., Sellami, L., & Hamida, A. B. , "Automatic classification and localization of prostate cancer using multi-parametric MRI/MRS," *Biomedical Signal Processing and Control,* vol. 31, pp. 189-198, 2017.

[68] Chapelle, O., Haffner, P., & Vapnik, V. N., "Support vector machines for histogram-based image classification," *IEEE transactions on Neural Networks,* vol. 10, no. 5, pp. 1055-1064, 1999.

[69] Ben-Hur, A., Ong, C. S., Sonnenburg, S., Schölkopf, B., & Rätsch, G., "Support vector machines and kernels for computational biolog," *PLoS computational biology,* vol. 4, no. 10, 2008.

[70] Montazer, G. A., & Giveki, D. , "An improved radial basis function neural network for object image retrieval," *Neurocomputing,* vol. 168, pp. 221-233, 2015.

[71] Dietterich, T. G., "Approximate statistical tests for comparing supervised classification learning algorithms," *Neural computation,* vol. 10, no. 7, pp. 1895-1923, 1998.